\documentclass[twocolumn,a4paper,pre,superscriptaddress,longbibliography]{revtex4-1}
\usepackage{tikz}
\usetikzlibrary{calc}
\usepackage{graphicx}
\usepackage{amssymb}
\usepackage{amsmath}
\usepackage{amsthm}

\usetikzlibrary{calc}
\usetikzlibrary{matrix}
\tikzstyle{commdiag}=[matrix of math nodes, row sep=3em, column sep=5.5em, text height=1.5ex, text depth=0.25ex,ampersand replacement=\&]
\tikzset{>=stealth}

\begin{document}

\graphicspath{ {../Olivier/riemann/} }

\def\R{\mathbb{R}}
\def\s{\boldsymbol{s}}
\def\x{\boldsymbol{x}}
\def\u{{\boldsymbol{u}}}
\def\v{\bar{\boldsymbol{v}}}
\def\w{\bar{\boldsymbol{w}}}
\def\dS{\hat{\s}}i
\def\bS{\bar{\s}}
\def\du{\hat{\boldsymbol{u}}}
\def\dv{\hat{\boldsymbol{v}}}
\def\dw{\hat{\boldsymbol{w}}}
\def\H{\mathbf{H}}
\def\tr{\mathrm{tr}}
\def\da{\hat{\boldsymbol{\alpha}}}
\def\db{\hat{\boldsymbol{\beta}}}
\def\ba{\bar{\boldsymbol{\alpha}}}
\def\bb{\bar{\boldsymbol{\beta}}}
\def\a{\boldsymbol{\alpha}}
\def\b{\boldsymbol{\beta}}
\def\f{\boldsymbol{f}}
\def\dt{\Delta t}
\newcommand{\kmnew}[1]{{\color{red}#1}}
\newcommand{\kmchange}[2]{{\color{red}#1} {\color{blue}[#2]}}

\newtheorem{theorem}{Theorem}
\newtheorem{corollary}[theorem]{Corollary}
\newtheorem{prop}[theorem]{Proposition}
\newtheorem{lem}[theorem]{Lemma}
\newtheorem{dfn}[theorem]{Definition}

\title{Symplectic integrators for spin systems}
\author{Robert I McLachlan}
\email{r.mclachlan@massey.ac.nz}
\affiliation{Institute of Fundamental Sciences, Massey University, Palmerston North, New Zealand}
\author{Klas Modin}
\email{klas.modin@chalmers.se}
\affiliation{Mathematical Sciences, Chalmers University of Technology, Gothenburg, Sweden}
\author{Olivier Verdier}
\email{olivier.verdier@math.umu.se}
\affiliation{Mathematics and Mathematical Statistics, Ume\aa~Universitet, SE--901 87 Ume\aa, Sweden}

\date{\today}

\begin{abstract}
We present a symplectic integrator, based on the implicit midpoint method, 
for classical spin systems where each spin is a unit vector in $\R^3$. 
Unlike splitting methods, it is defined for all Hamiltonians and is $O(3)$-equivariant, i.e., coordinate-independent.
It is a rare example of a generating function for symplectic maps of a noncanonical phase space. 
It yields a new integrable discretization of the spinning top.
\end{abstract}

\maketitle
 
Symplectic integrators for the computer simulation of Hamiltonian dynamics are widely used in computational physics~\cite{leimkuhler2004simulating,hairer2006geometric}. 
For canonical Hamiltonian systems, with phase space $\R^{2N}$ and canonical coordinates $(q^i,p_i)$, simple and effective symplectic integrators  are known. 
For noncanonical systems, like spin systems with phase space $(S^2)^N$, some symplectic integrators are known.
These are, however, either (i) based on local coordinates and not rotationally invariant, (ii) defined only for special Hamiltonians, or (iii) excessively complicated with many auxiliary variables.
Here we solve the computational physics problem of providing a globally-defined, rotationally invariant, minimal-variable symplectic integrator for general spin systems. 
The method is surprisingly simple and  depends only on the vector field of the system at hand. 
It is a rare example of a generating function for symplectic maps on a noncanonical phase space: a noncanonical analogue of  the Poincar\'e generating function of classical mechanics. 
The method produces new discrete-time physical models, such as a new completely integrable discrete spinning top, and unveils new directions for symplectic integrators, discrete physics, and symplectic geometry.

Classical spin systems are a class of noncanonical Hamiltonian systems with phase space $(S^2)^N$ and symplectic form the sum of the standard area elements on each sphere. 
If the spheres are realized as $\|\s_i\|^2 = 1$, $\s_i\in\R^3$, and $H$ is the Hamiltonian on $(S^2)^N$ arbitrarily extended to $(\R^3)^N$, the equations of motion take the form
\begin{equation}
\label{eq:spin}
\dot{\s_i} = \underbrace{\s_i\times\nabla_{\s_i}H(\s_1,\dots,\s_N)}_{\displaystyle\f_i(\s_1,\ldots,\s_N)} .
\end{equation}
Spin systems include the classical limit of quantum (e.g. Heisenberg) spin chains, (discretizations of) the Landau--Lifshitz equation of micromagnetics \cite{lakshmanan2011fascinating}, and point vortices on the sphere \cite{pekarsky1998point}. 
Single-spin systems include the reduced motion of a spinning top (free rigid body) \cite{marsden1999introduction} and the motion of a particle advected by an incompressible 2D fluid on a sphere. 

Our main result is a new integrator for~\eqref{eq:spin} given by
\begin{equation}\label{eq:main_algorithm}
\begin{split}
	\frac{\s_{i,n+1}-\s_{i,n}}{\Delta t} &= \f_i(\boldsymbol{u}_1,\dots,\boldsymbol{u}_N), \\
\boldsymbol{u}_i &:= \frac{\s_{i,n} + \s_{i,n+1}}{\|\s_{i,n} + \s_{i,n+1}\|}.
\end{split}	
\end{equation}
This \emph{spherical midpoint method} is globally defined and preserves many structural properties of the exact flow.
Before explaining these properties we review symplectic integrators for canonical and noncanonical systems. 

Symplectic integrators for canonical Hamiltonian systems fall into two main classes: explicit methods, based on splitting the Hamiltonian into integrable terms and composing their flows, and implicit methods, typically based on generating functions. 
(Discrete Lagrangians can generate both types of method.) 
The \emph{leapfrog} or \emph{St\"ormer--Verlet method}, almost universally used in molecular dynamics, is an example of an explicit method, whereas the \emph{classical midpoint method}
\begin{equation}\label{eq:midpoint_method}
	\frac{\boldsymbol{z}_{n+1}- \boldsymbol{z}_n}{\Delta t} = \boldsymbol{F}\left(\frac{\boldsymbol{z}_n + \boldsymbol{z}_{n+1}}{2}\right),
\end{equation}
for $\dot{\boldsymbol{z}} = \boldsymbol{F}(\boldsymbol{z})$ with $\boldsymbol{z} \in \R^{2N}$, is an example of an implicit method.
The classical midpoint method~\eqref{eq:midpoint_method} has a number of striking features:
(i) it is defined for {\em all} Hamiltonians in a uniform way (splitting methods are only defined for separable Hamiltonians);
(ii) it conserves quadratic invariants;
(iii) it is equivariant with respect to all affine maps of phase space (that is, it is intrinsically defined on the affine phase space and does not depend on the choice of affine coordinates; it does not require canonical coordinates);
(iv) it preserves all affine symmetries and foliations;
(v) it is unconditionally stable for linear systems, which confers somewhat improved stability for nonlinear systems;
(vi) it is self-adjoint under $t\to -t$ and preserves all affine time-reversing symmetries;
(vii) it is symplectic for all constant symplectic structures, Poisson for all systems with constant Poisson structure, and pre-symplectic for all systems with constant pre-symplectic structure \cite{verdiersymplectic};
(viii) it is a Runge--Kutta method, which allows the application of an extensive body of numerical analysis including forward and backward error analysis and the construction of the modified (numerical) Hamiltonian; and
(ix) it is a symplectic map associated with the Poincar\'e generating function \cite[vol. III, \S319]{poincare}
\begin{equation}\label{eq:poincare_generating_fun}
	\Omega(\boldsymbol{z}_{n+1} - \boldsymbol{z}_n) =\nabla G\left(\frac{\boldsymbol{z}_n + \boldsymbol{z}_{n+1}}{2}\right),\; \Omega = \begin{pmatrix}0 & I \cr -I & 0 \end{pmatrix}
\end{equation}
with the generating function $G$ chosen to be the product of the time step and the Hamiltonian.
Because of these properties, the classical midpoint method has a claim to be the `natural' discrete time analogue of Hamiltonian vector fields on symplectic vector spaces; it is indeed extensively used in computational physics~\cite{mandziuk1995resonance,
preto2009post,
schafer2008implicit,
dubinkina2007statistical,
hellstrom2009satellite,
brown2006midpoint,
zhong2010global,
wu2010symplectic}.

Symplectic integrators are known for some noncanonical Hamiltonian systems. 
The most commonly used approach is splitting \cite{mclachlan1993explicit,touma1994lie,frank1997geometric,dullweber1997symplectic,omelyan2001algorithm,meineke2005oopse}; as in the canonical case, this requires the Hamiltonian to have a special structure and the splitting to be designed by hand. Current general-purpose methods for Lie--Poisson systems for general Hamiltonians tend to be complicated and involve implicit equations involving infinite series of Lie brackets \cite{zhong1988lie,channell1991integrators,marsden1999discrete} and extra variables \cite{reich1994momentum,mclachlan1995equivariant,mclachlan2013collective}. 
The classical midpoint method itself is not symplectic when applied to spin systems (\ref{eq:spin}); this was noted already in the single spin case in \cite{austin1993almost}. Despite this, it has been used in some applications to spin systems, for its other favorable properties \cite{d2006midpoint,mentink2010stable}:
it is $O(3)$-equivariant (it commutes with rotations and reflections; its dynamics are independent of the choice of coordinates), preserves the spin lengths $\|\s_i\|$, and is linearly stable for all $\Delta t$.
More generally, there is a lack of generating functions---the most fundamental tool in classical mechanics---for noncanonical phase spaces. 

We now discuss properties of the new method~\eqref{eq:main_algorithm}.
First, a key observation: our method coincides with the classical midpoint method 
applied to the vector field 
\begin{equation*}
	\boldsymbol{g}_{i}(\s_1,\ldots,\s_N) := \f_i\left( \frac{\s_1}{\|\s_1\|},\ldots,\frac{\s_N}{\|\s_N\|} \right).
\end{equation*}
This immediately implies several properties: 
(i) it preserves the spin lengths $\|\s_i \|$;
(ii) it is $O(3)$--equivariant;
(iii) it is second-order accurate;
(iv) it is self-adjoint; and
(v) it preserves arbitrary linear symmetries, arbitrary linear integrals, and single-spin homogeneous quadratic integrals $\s_i^T \mathbf{A} \s_i$.
Symplecticity is not, however, an immediate result, since the symplectic structure of $(S^2)^N \subset \R^{3N}$ is nonlinear.
Nevertheless, the method is symplectic.
There are two ways to show this: a direct proof incorporating new techniques based on ray-constant Hamiltonians and linearity of the Lie--Poisson structure, and a geometric proof based on the extended Hopf map and realization of the spherical midpoint method as a \emph{collective symplectic method}~\cite{mclachlan2013collective}.
Both proofs are given in~\cite{mclachlan2013spherical}.
Because of its symplecticity, the spherical midpoint method can be interpreted as a generating function on $(S^2)^N$, 
analogous to the Poincar\'e generating function~\eqref{eq:poincare_generating_fun} on $\R^{2N}$.

Let us briefly consider single-spin systems, i.e., $N=1$.
If $H$ is of the form $H(\s) = \sum_{j=1}^3 s_j^2/(2 I_j)$ with $I_j > 0$ (spinning top), then the spherical midpoint method exactly conserves $H$ (since it is a homogeneous quadratic invariant).
Since the method is symplectic and also conserves the total angular momentum $\| \s \|$, the corresponding discrete dynamical system $\s_n \mapsto \s_{n+1}$ is completely integrable. This situation may be compared to
the Moser--Veselov discretization~\cite{moser1991discrete} of the spinning top~\cite{hairer2006preprocessed}. This hugely influential discretization of  tops, and more generally of any Lie-Poisson system on the dual of the Lie algebra $\mathfrak{g}$ of a Lie group $G$, suspends the continuous Lagrangian to $TG$ and constructs a discrete Lagrangian on $G\times G$ by embedding $G$ in a linear space of matrices and discretize velocities $\dot{\mathbf{Q}}$ by $(\mathbf{Q}_{n+1}-\mathbf{Q}_n)/(\Delta t)$. 
The final algorithm requires solving nonlinear equations in $G$ ($SO(3)$ for the spinning top, $SO(3)^N$ for spin systems) and is closely related to the \textsc{rattle} method of molecular dynamics \cite{leimkuhler2004simulating,mclachlan1995equivariant}. 
Remarkably, the Moser--Veselov discretization is completely integrable for many systems including the spinning top. 
It also describes the eigenstates of certain quantum spin chains. 
Its relationship to other integrable discrete physics models, that typically do not arise from a simple variational principle, is not clear. 
In this context it is striking that the spherical midpoint method gives a {\em different} integrable discrete version of the spinning top, arising not from a variational principle but from a standard numerical integrator, related to the fundamental Poincar\'e generating function for canonical systems.

\begin{figure}[htp]
	\centering
	\begin{tikzpicture}
		\node[anchor=south west, inner sep=0] (image) at (0,0) {\includegraphics[width=0.35\textwidth]{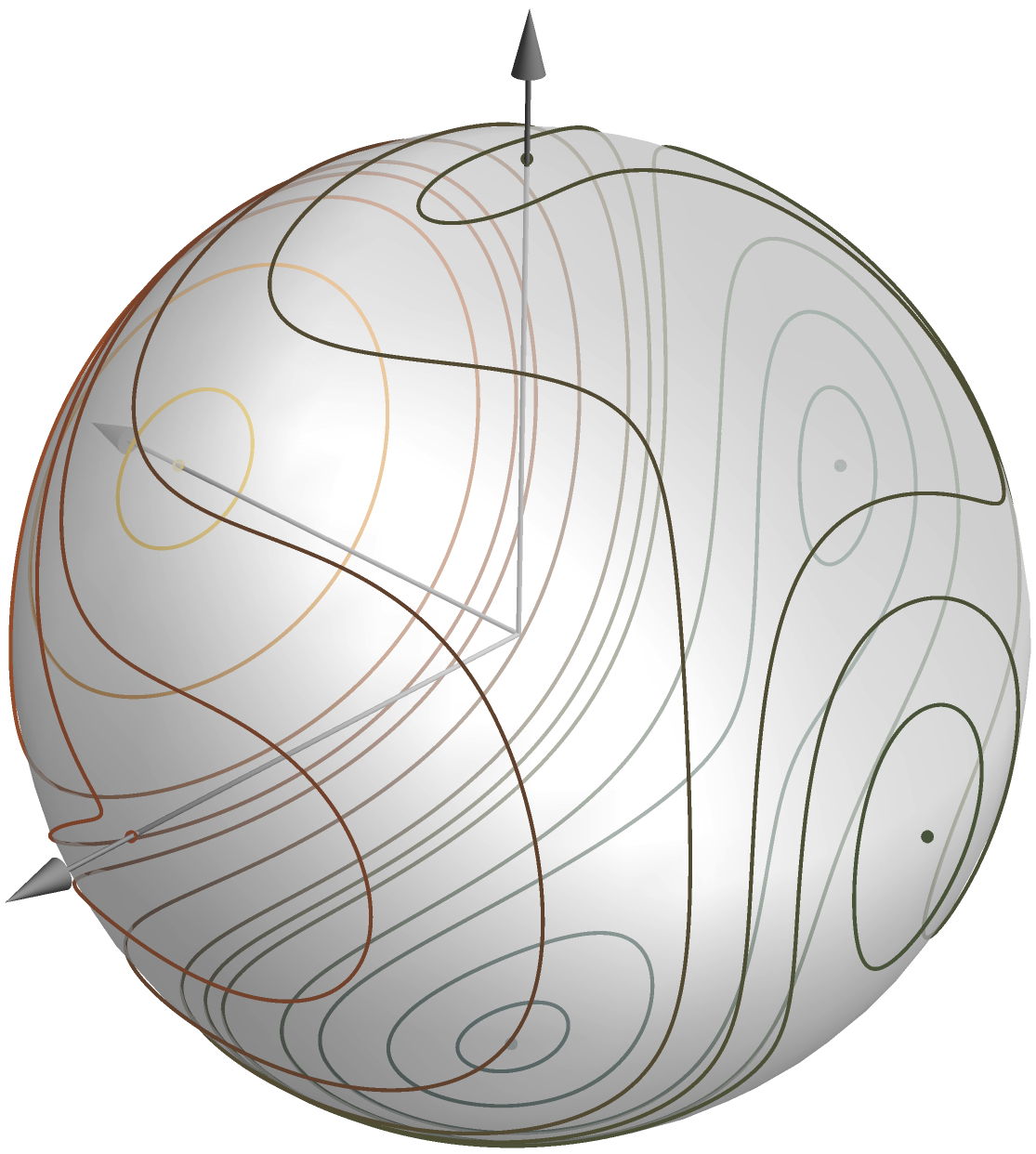}};
		\begin{scope}[x={(image.south east)},y={(image.north west)}]
			\coordinate (x) at (0.025,0.66) {};
			\coordinate (y) at (-0.01,0.27) {};
			\coordinate (z) at (0.45,0.97) {};
			\node[below=-1ex] at (x) {$s_1$};
			\node[below=-1ex] at (y) {$s_2$};
			\node[below=-1ex] at (z) {$s_3$};
		\end{scope}
	\end{tikzpicture}
	\caption{
		The sphere $\|\s\|=1$ is shown together with the phase portrait of the single-spin system with Hamiltonian~\eqref{eq:Ham_nonsymrb}. ? steps with $\Delta t = ?$ are shown for different initial conditions, resulting in 13 periodic orbits and 6 equilibria.
	}
	\label{nonsymrb_phasediagram_fig}

	\begin{tikzpicture}
		\node[anchor=south west, inner sep=0] (image) at (0,0) {\includegraphics[width=0.35\textwidth]{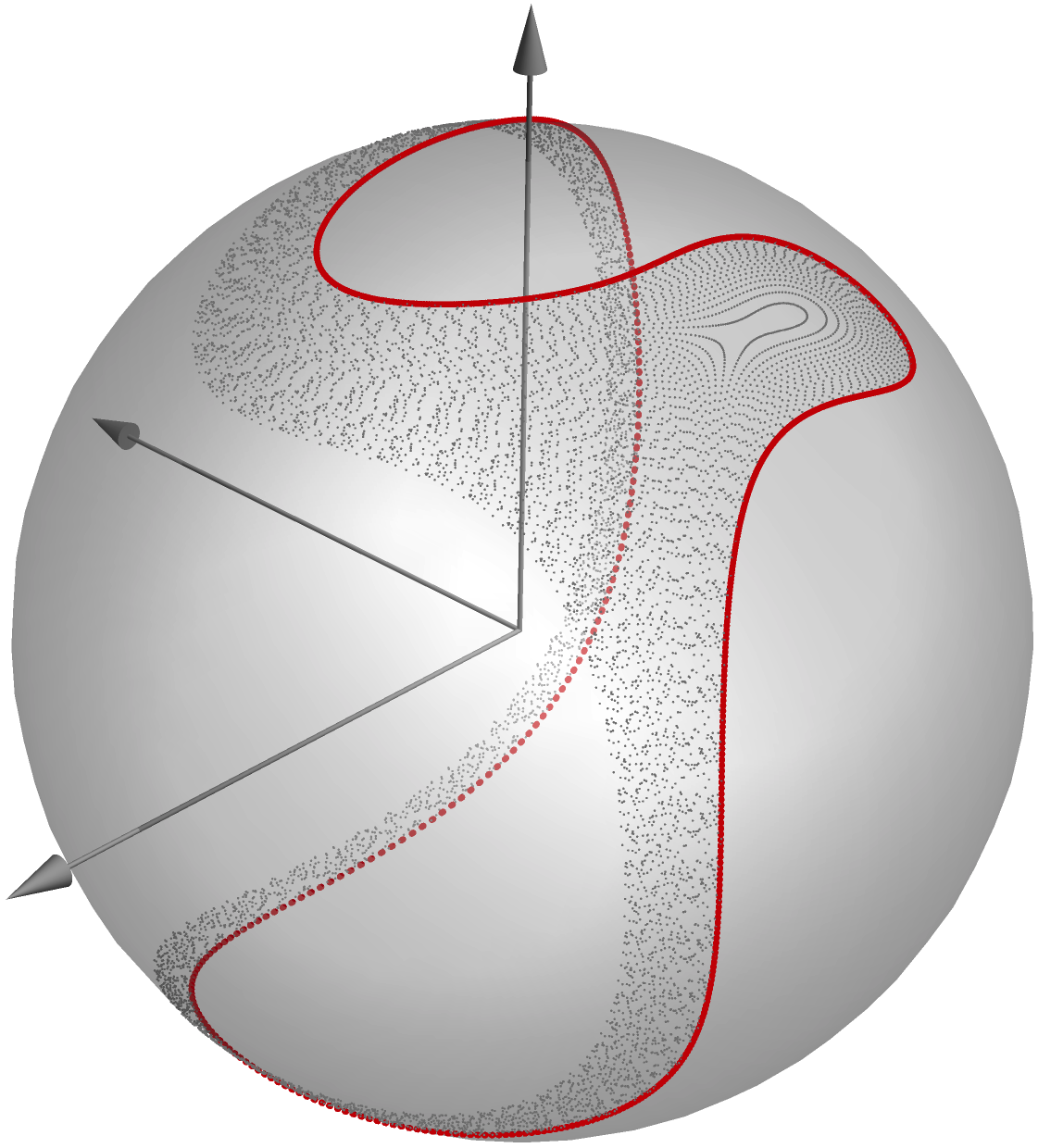}};
		\begin{scope}[x={(image.south east)},y={(image.north west)}]
			\coordinate (x) at (0.025,0.66) {};
			\coordinate (y) at (-0.01,0.27) {};
			\coordinate (z) at (0.45,0.97) {};
			\node[below=-1ex] at (x) {$s_1$};
			\node[below=-1ex] at (y) {$s_2$};
			\node[below=-1ex] at (z) {$s_3$};
		\end{scope}
	\end{tikzpicture}
	\caption{
		Discrete trajectories for the single-spin system with Hamiltonian~\eqref{eq:Ham_nonsymrb} obtained using the classical midpoint method (dots) and the spherical midpoint method (thick line). 
		The initial condition is  $\s_0 = (0,0.7248,-0.6889)$.
		The time step is $\Delta t=0.5$.
		The trajectory is periodic for the symplectic spherical midpoint method (correct behavior), but non-periodic for the nonsymplectic classical midpoint method (incorrect behavior).
	}
	\label{nonsymrb_drift_fig}
\end{figure}

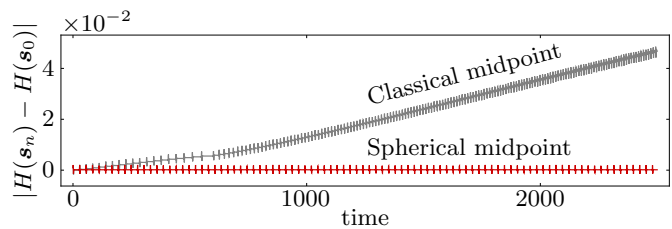
\begin{figure}[htp]
	\begin{tikzpicture}
		\node[anchor=south west, inner sep=0] (image) at (0,0) {\includegraphics[width=0.45\textwidth]{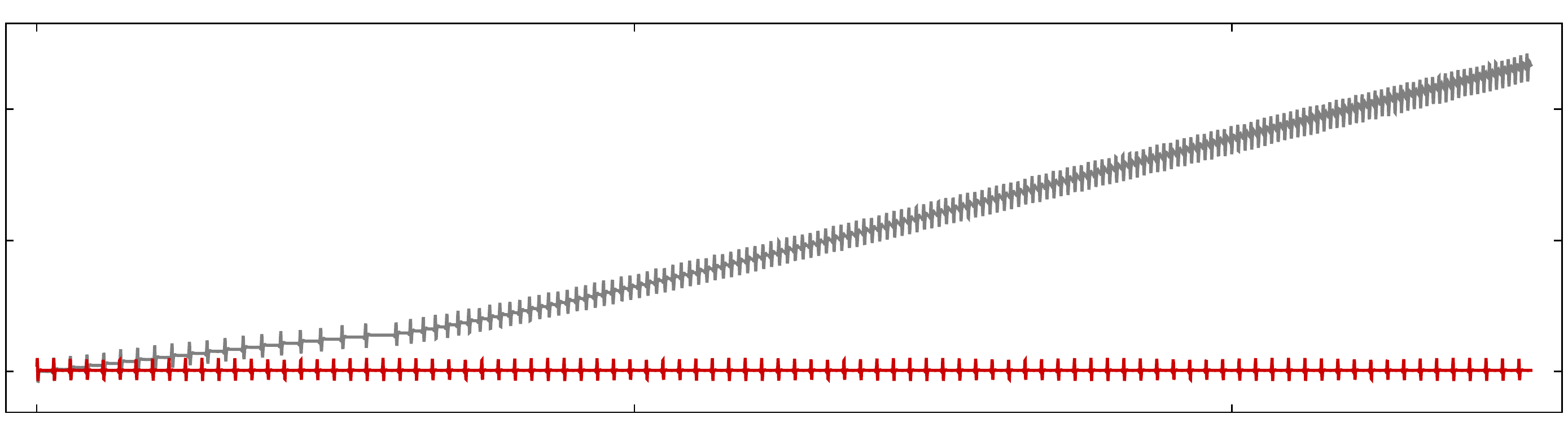}};
		\begin{scope}[x={(image.south east)},y={(image.north west)}]
			\coordinate (x1) at (0.025,0.02) {};
			\coordinate (x2) at (0.403,0.02) {};
			\coordinate (x3) at (0.785,0.02) {};
			\node[below=-1ex] at (x1) {\small$0$};
			\node[below=-1ex] at (x2) {\small$1000$};
			\node[below=-1ex] at (x3) {\small$2000$};
			\coordinate (y1) at (0,0.75) {};
			\coordinate (y2) at (0,0.452) {};
			\coordinate (y3) at (0,0.152) {};
			\node[left=-0.5ex] at (y1) {\small$4$};
			\node[left=-0.5ex] at (y2) {\small$2$};
			\node[left=-0.5ex] at (y3) {\small$0$};
			\coordinate (exp) at (0,1.02) {};
			\node[right=-0.5ex] at (exp) {\small$\times 10^{-2}$};	
			\coordinate (label1) at (0.67,0.6) {};
			\coordinate (label2) at (0.67,0.16) {};
			\node[above=0ex,rotate=14] at (label1) {\small Classical midpoint};
			\node[above=0ex] at (label2) {\small Spherical midpoint};
			\coordinate (xlabel) at (0.5,0) {};
			\coordinate (ylabel) at (-0.02,0.5) {};
			\node[below=0.5ex] at (xlabel) {\small time};
			\node[above=0ex,rotate=90] at (ylabel) {\small $|H(\s_n)-H(\s_0)|$};
		\end{scope}
	\end{tikzpicture}
	\caption{
		Energy error versus time for the classical and spherical midpoint methods applied to the single-spin system with Hamiltonian~\eqref{eq:Ham_nonsymrb}.
		The initial condition is  $\s_0 = (0,0.7248,-0.6889)$.
		The time step is $\Delta t=0.5$.
		The energy drifts for the classical midpoint method, but remains bounded for the spherical midpoint method.
	}
	\label{fig_nonsymrb_energy}
\end{figure}

Two brief examples illustrate the behavior of the method on an integrable and a nonintegrable single-spin system.
The first has Hamiltonian 
\begin{equation}\label{eq:Ham_nonsymrb}
H(\s) = \frac{1}{2}\sum_{j=1}^3 \frac{1}{I_j} (s_j^2 + \frac{2}{3} s_j^3),\quad I=(1,2,4),
\end{equation}
and is a nonlinear perturbation of a spinning top. 
Like the spinning top, all orbits are periodic, as shown by the phase diagram in Fig.~\ref{nonsymrb_phasediagram_fig}. 
Computed trajectories for the spherical and classical midpoint methods are shown in Fig.~\ref{nonsymrb_drift_fig}: 
trajectories lie on smooth curves for the spherical midpoint method but not for the classical midpoint method.
Energy errors are shown in Fig.~\ref{fig_nonsymrb_energy}: the energy error is bounded for the spherical midpoint method but grow in time for the classical midpoint method.
These results are consistent with the symplecticity (or lack thereof) of the methods.

\begin{figure}[htp]
	\centering
	\begin{tikzpicture}
		\node[anchor=south west, inner sep=0] (image) at (0,0) {\includegraphics[width=0.4\textwidth]{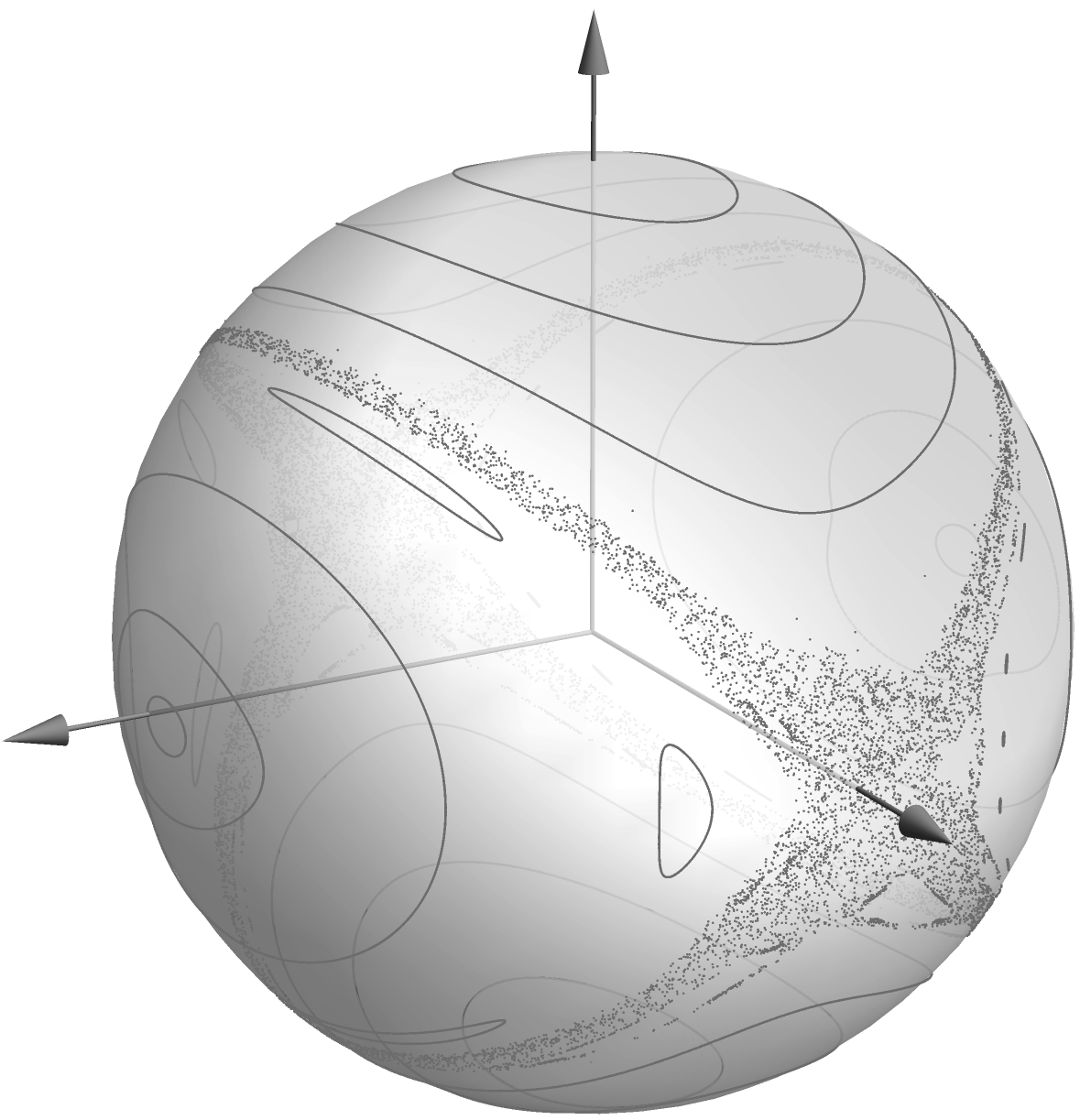}};
		\begin{scope}[x={(image.south east)},y={(image.north west)}]
			\coordinate (x) at (0.025,0.4) {};
			\coordinate (y) at (0.97,0.23) {};
			\coordinate (z) at (0.5,0.97) {};
			\node[below=-1ex] at (x) {$s_1$};
			\node[below=-1ex] at (y) {$s_2$};
			\node[below=-1ex] at (z) {$s_3$};
		\end{scope}
	\end{tikzpicture}
	\caption{
		Poincar\'e section (one-period map) of the periodically forced spinning top system with Hamiltonian~\eqref{eq:forced_rb_Ham} with $\varepsilon=0.07$, approximated by the spherical midpoint method with 20 time steps per period.
	}
	\label{forced_rb_fig}
\end{figure}
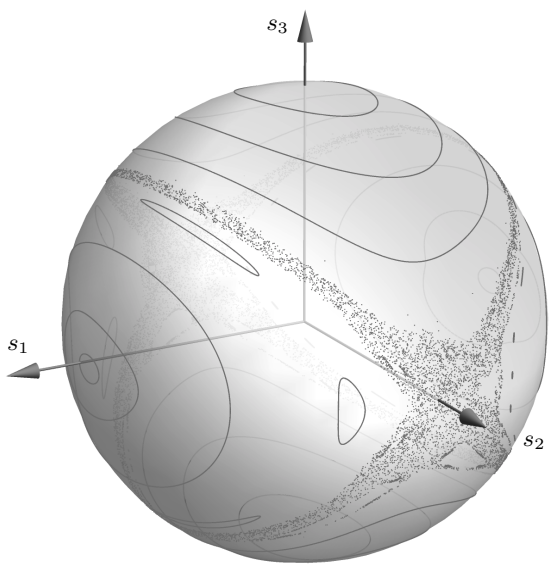

Our second example is a periodically forced spinning top with Hamiltonian
\begin{equation}\label{eq:forced_rb_Ham}
	H(\s,t) = \frac{1}{2}\sum_{j=1}^3 \frac{s_j^2}{I_j} + \varepsilon\sin(t)s_3,\quad I=(1,{\textstyle\frac{4}{3}},2).
\end{equation}
The phase portrait of the 1-period (Poincar\'e) map obtained using 
the spherical midpoint method with time-step length $2\pi/k$, $k=20$, is shown in Fig.~\ref{forced_rb_fig}, and illlustrates the breakup of heteroclinic and periodic orbits, and a transition to chaos, typical of this class of systems.

The spherical midpoint method is implicit. Implicit methods are most often used on stiff systems, like reaction--diffusion and fluid systems, that contain widely-varying timescales. In these cases sophisticated solvers are needed. For the present case, and in other applications of the classical midpoint method~\cite{mandziuk1995resonance,
preto2009post,
schafer2008implicit,
dubinkina2007statistical,
hellstrom2009satellite,
brown2006midpoint,
zhong2010global,
wu2010symplectic}, the fixed-point iteration
\begin{equation*}
\boldsymbol{z}_{n+1}^{(0)}=\boldsymbol{z}_n,\ \boldsymbol{z}_{n+1}^{(k+1)} =\boldsymbol{z}_n + \Delta t \boldsymbol{F}((\boldsymbol{z}_n +\boldsymbol{z}_{n+1}^{(k)})/2),\ k\ge 0
\end{equation*}
applied to (\ref{eq:midpoint_method}) is often sufficient, terminating when $\|\boldsymbol{z}_{n+1}^{(k+1)}-\boldsymbol{z}_{n+1}^{(k)}\|$ is less than some chosen tolerance. For typical time steps this can take 5--10 iterations. However, more sophisticated iterations are possible \cite{mclachlan2007new} and can lead to implementations that use 2 evaluations of the vector field (here, $\boldsymbol{F}$) per time step. Special termination criteria can improve the propagation of roundoff error \cite{hairer2008achieving}.

The method \eqref{eq:main_algorithm} is the first equivariant symplectic integrator for spin systems that does not contain auxiliary variables. 
Since the spin lengths $\|\s_i\|$ are preserved, effectively the method requires the solution of $2N$ nonlinear equations per step, which is just the dimension of the phase space. 
Not only is \eqref{eq:main_algorithm} very simple, it does not depend on or even require a {\em formula} for $\boldsymbol{f}$: in some applications, for example to the advection of particles by an incompressible fluid on the sphere, $\boldsymbol{f}$ may be provided by a `black box' which may involve experimental data, local or global interpolation, or the output of a separate CFD code. 
Further details and properties of the method, including its connection to collective symplectic integrators and Riemannian integrators, and numerical experiments, may be found in \cite{mclachlan2013spherical}. 
The supplementary material contains animations illustrating the method applied to systems with multiple spins, including the Heisenberg spin chain ($N=100$) and point vortices on the sphere ($N=8$, $12$).

The method extends in the obvious way to arbitrary spin--liquid systems \cite{omelyan2001algorithm} with phase space $(T^*\R^3\times S^2)^N$. 
The method can be generalized to yield symplectic integrators for Nambu-type systems $\dot{\s_i} = \nabla C_i(\s_i)\times\nabla_i H(\s_1,\dots,\s_N)$, where each $C_i$ is a homogeneous quadratic \cite{mclachlan2013spherical}; the phase space is a product of classical conic sections. 
It is an open question as to for which symplectic manifolds such an integrator (or generating function) exists.

\bibliography{smp}

\end{document}